\documentclass[10pt,conference]{IEEEtran}
\IEEEoverridecommandlockouts

\usepackage{cite}
\usepackage{caption}

\usepackage{amsmath,amssymb,amsfonts, wasysym}
\usepackage{algorithmic}
\usepackage{graphicx}
\usepackage{textcomp}
\usepackage{xcolor}
\usepackage{enumitem}
\usepackage{footnote, url}

\usepackage{colortbl}
\usepackage{hhline}
\usepackage{cancel}
\usepackage{makecell}
\usepackage{multirow}
\usepackage{rotating}
\usepackage{hyperref}

\usepackage{colortbl}
\usepackage{booktabs}

\usepackage{xcolor}

\definecolor{generalChallenges}{RGB}{243, 232, 238}
\definecolor{devopsSpecificChallenges}{RGB}{255, 204, 204}
\definecolor{teachingMethods}{RGB}{114, 155, 121}

\def\BibTeX{{\rm B\kern-.05em{\sc i\kern-.025em b}\kern-.08em
    T\kern-.1667em\lower.7ex\hbox{E}\kern-.125emX}}
\begin{document}

\title{Overcoming Challenges in DevOps Education\\ through Teaching Methods}

\author{\IEEEauthorblockN{Samuel Ferino\textsuperscript{1}, Marcelo Fernandes\textsuperscript{1,2}, Elder Cirilo\textsuperscript{3}, Lucas Agnez\textsuperscript{1},\\ Bruno Batista\textsuperscript{1}, Uirá Kulesza\textsuperscript{1}, Eduardo Aranha\textsuperscript{1}, Christoph Treude\textsuperscript{4}}
\IEEEauthorblockA{\vspace*{-1em}\\\textsuperscript{1}\textit{Federal University of Rio Grande do Norte, Brazil}; \textsuperscript{2}\textit{Federal Institute of Rio Grande do Norte, Brazil}\\
\textsuperscript{3}\textit{Federal University of São João del Rei, Brazil}; \textsuperscript{4}\textit{University of Melbourne, Australia}\\
\vspace*{-1em}\\samuellucas97@ufrn.edu.br, marcelo.fernandes@ifrn.edu.br, elder@ufsj.edu.br, lucasagnez@hotmail.com\\
brunokaike2014@gmail.com, \{uira,eduardoaranha\}@dimap.ufrn.br, christoph.treude@unimelb.edu.au}
}

\maketitle

\begin{abstract}
DevOps is a set of practices that deals with coordination between development and operation teams and ensures rapid and reliable new software releases that are essential in industry. DevOps education assumes the vital task of preparing new professionals in these practices using appropriate teaching methods. However, there are insufficient studies investigating teaching methods in DevOps. We performed an analysis based on interviews to identify teaching methods and their relationship with DevOps educational challenges.  Our findings show that \textsl{project-based learning} and \textsl{collaborative learning} are emerging as the most relevant teaching methods.
\end{abstract}

\begin{IEEEkeywords}
DevOps, teaching methods, challenges, mixed methods\end{IEEEkeywords}

\section{Introduction}

DevOps (Development and Operations) appears as a natural evolution
of the Agile movement \cite{almeida:2022, bobrov:2020}. It aims to mitigate conflicts, approximate responsibilities, and improve the communication of Development and Operation teams
\cite{ieee:2021, bahadori:2018, rodrigues:2017, ebert:2016}. This Agile perspective was essential for organizations to remain competitive in the technological business, meeting the need to build fast, resilient, and secure deliverables that involve distributed systems at scale. Recent studies show that the evolution of DevOps practices in industry organizations aligns with business success \cite{forsgren:2018}. Since 2009, there have been various examples of large organizations adopting DevOps, including Facebook, Yahoo, Flickr, Netflix, and Etsy \cite{rong:2017, erich:2017, bang:2013}. Using DevOps practices, companies such as Amazon and Netflix deploy thousands of times a day \cite{agarwal:2018}.

In many IT sectors, mastering DevOps practices has become a necessary skill \cite{wiesche:2018, spinellis:2016}. There is a high demand for DevOps practitioners, with many related job postings \cite{pang:2020}. However, there is a shortage of qualified professionals to meet demand \cite{ferino:2021, krusche:2014}. Because of that, the industry has a growing desire to have graduates educated in the use of DevOps approaches and tools, better preparing students to apply DevOps practices in real projects \cite{jennings:2019, bobrov:2019}.

DevOps teaching has its challenges \cite{christensen:2016}.  From an educational point of view, it is difficult to set up the DevOps approach from scratch \cite{ferino:2021, bobrov:2019}. The challenge of creating an effective approach intensifies when considering students who do not have the necessary background to continuously evolve in learning. For example, students who have not worked in industry or on projects of substantial size and complexity have difficulty understanding DevOps \cite{jones:2018, perez:2021}. More, DevOps research with an education focus is required to investigate these aspects.

Investing in DevOps teaching is a meaningful way to mitigate industry demand. From an educational perspective, the role of the educator is to guide students throughout the learning process by applying the proper teaching methods and providing a suitable learning environment \cite{kilamo:2012}.

Although there is no predominant approach, some approaches are more effective than others, depending on the teaching context \cite{marques:2014}. \textsl{Project-based learning} teaching practice, for example, has been widely accepted as an important part of Software Engineering (SE) Education \cite{bai:2018, chatley:2017, moreno:2016}. Educators and practitioners agree that the \textsl{lecture} approach (traditional teaching method) is not enough to teach SE \cite{chen:2009}. Inactiveness of students, tiring lectures, one-way communication, and fast forgetting are the main disadvantages of this method \cite{sadeghi:2014, norouzi:2011}.

In this sense, approaching teaching from different perspectives comes from the idea that all people learn differently. Therefore, teaching must be considered as a changing variable. The way of approaching diversity in students also impacts the variety of teaching \cite{chen:2009}. Another approach available to educators is to combine different teaching methods and approaches in a unique course. This idea is not new, as can be seen in the combination of theoretical (\textsl{lectures}) and practical (\textsl{lab}) classes. However, the challenge is to justify all teaching methods to fit into one course \cite{kuhrmann:2012}.

Ferino \textit{et al.} \cite{ferino:2021} presented an initial study of teaching methods in DevOps Education. They analyzed 18 papers published up to 2019 from a systematic literature review of DevOps teaching experience. This study started the research on teaching methods in the DevOps area.  However, more research is required to validate and improve the knowledge of DevOps teaching methods by educators, as this is a preliminary study. 

This work aims to investigate the teaching methods that are being used in DevOps courses and how these methods address the challenges found in the DevOps Education area. To achieve this goal, we seek to answer the following research questions (RQs):

{\renewcommand\labelitemi{}
\begin{itemize}[leftmargin=*]
\item  \textbf{RQ1}. \textit{What are the teaching methods and approaches used in DevOps courses?}
\item  \textbf{RQ2}. \textit{How can teaching methods address the challenges of DevOps courses?}
\end{itemize}
}

These questions are answered by analyzing the data resulting from the study of Fernandes \textit{et al.} \cite{fernandes:2022}, who interviewed 14 DevOps educators. That work analyzed challenges and recommendations in DevOps Education. Our paper focuses on identifying teaching methods and linking them to DevOps education challenges. As a result, this work provides the following contributions:

 \begin{itemize}
\item Identification of 18 interview-derived teaching methods that can be implemented in DevOps courses to improve student learning. Our study identified 7 new teaching methods compared to a previous literature study \cite{ferino:2021}: \textsl{mentoring},  \textsl{review session},  \textsl{personalized learning},  \textsl{seminar},  \textsl{comprehensive distance learning}, \textsl{feedback session}, and \textsl{example-based learning}. 

\item A comprehensive set of 44 links between teaching methods and DevOps education challenges. Each link has a specific recommendation on how to use the teaching method to tackle the challenges. These links involve many teaching methods, different from the literature \cite{ferino:2021} that focuses on the links between challenges and the \textsl{educational support tool} teaching method.

\end{itemize}

The remainder of this paper is structured as follows.  Section~\ref{sec:relatedWork} presents related work. Then, Section~\ref{sec:methodology} overviews the methodology used in this work. The report and discussion of teaching methods and related challenges occur in Sections~\ref{sec:results}  and~\ref{sec:discussion}.
Section~\ref{sec:threatsToValidity} reviews the threats to the validity of this study. Final remarks about this work and discussions about research opportunities are presented in Section~\ref{sec:conclusion}.

\section{Related Work}   \label{sec:relatedWork}

Ferino \textit{et al.} \cite{ferino:2021} present an initial study of teaching methods found in the literature, showing ways to help and mitigate challenges found in DevOps teaching. It relates to this work as it also looks for DevOps teaching methods, presenting 13 teaching methods. They investigated the \textsl{educational support tool} teaching method in more depth, focusing on linking it to educational DevOps challenges. Conversely, our study presents  a total of 18 teaching methods identified analyzing DevOps educators' interviews. Additionally, we have validated the results, employing internal validation by member checking and external comparison with the study by Ferino \textit{et al.} \cite{ferino:2021}.

Grotta \textit{et al.} \cite{grotta:2022} present a systematic literature review that investigates the performance of students with Information System (IS) teaching applied in DevOps. They show a correlation between teaching DevOps and professional and academic development. It also shows that DevOps is strongly related to \textsl{project-based learning}. That work is related to ours because it explores teaching methods applied to DevOps. However, that work focused on linking DevOps to the Information Systems area, while our work seeks to identify teaching methods and link them to the challenges found in DevOps courses.

Fernandes \textit{et al.} \cite{fernandes:2022} present an analysis of interviews with DevOps educators from around the world, collecting challenges and recommendations from existing DevOps courses according to their experiences. This work is related to ours, as it investigates challenges in DevOps teaching. Our work differs from it, as it focuses on identifying teaching methods and relating them to these challenges.

\section{Methodology} \label{sec:methodology}

This section presents the methodology of this work, which is organized in two stages: (i) identifying the teaching methods used in DevOps courses; and (ii) linking these teaching methods to challenges. The basis for performing these analyzes is the data set produced by the study of Fernandes \textit{et al.} \cite{fernandes:2022}, an interview study conducted with 14 DevOps educators. These educators have many years of experience teaching computing courses. During the interviews, they shared their experiences teaching DevOps courses. All these courses have DevOps as the main focus and address different DevOps aspects  \cite{leite:2019} such as \textit{Runtime} and \textit{Delivery}. Our work identifies the teaching methods based on their interview transcripts and links them to the challenges highlighted by Fernandes \textit{et al.}

In every analysis for both stages, two people were involved, the main researcher and a reviewer researcher. Initially, the main researcher interprets the interview data and records observations and comments. The supporting researcher then reviews the records and discusses possible conflicting opinions with the main researcher. 
At this point, both researchers have to explain their reflections. A third researcher resolves the conflict if they do not achieve a common understanding. To help the researcher’s work and to store the analyses and results, we used \textit{Google Sheets}. We mention the interviewees using I terminology (I1, I2, ..., I14), where I10 refers to the tenth interview, for example.

\subsection{Identifying Teaching Methods}

The first stage of this work is to identify teaching methods based on the 14 interview transcripts. We are adopting Westwood \textit{et al.}'s \cite{westwood:2008} \textbf{teaching method definition}: \textit{a set of principles, procedures, or strategies to be implemented by teachers to promote desired learning in students} \cite{bjork:2017, liu:2007}. The following subsections provide more details on each step of this stage.

\subsubsection{Data Extraction}

The analysis starts by gathering mentions of teaching methods in the interviews' transcripts. For instance, consider the following interviewee's comment: \textit{``... I organize the students into teams of four or six per group, and then we work together. That's good also because they may be working in a team. They learn how to work in a team and team building..."} (I11). The mention of ``\textit{They learn how to work in a team and team building}" is an example of the teaching method \textsl{collaborative learning} (definition in Table \ref{tab:teachingMethodsDefinition}). 
Furthermore, 
there are direct mentions of teaching methods, as in \textit{``\underline{I use} not only \textbf{PBL}, but also \textbf{flipped classroom}''} (I7). 
Interviewee I7 cites two teaching methods used in her DevOps course: \textsl{problem-based learning} (PBL) and \textsl{flipped classroom}.

\subsubsection{Data Analysis}

Based on the extracted data, we performed open coding through the identification and categorization of teaching methods. It is important to use codification, as teaching methods may not be explicitly mentioned. For example, the following text snippet refers to the \textsl{feedback session} teaching method: \textit{``... I'm always \underline{taking almost an hour} for students to \underline{give me their feedback} ..."} (I12). We also observed whether the interviewees mentioned the use of teaching methods in their DevOps courses. Sometimes, we analyzed other parts of the interview transcripts to clarify the adoption of the methods. In this sense, sometimes it is not clear if the educator adopted \textsl{problem-based learning} during their course, as in the following quote: \textit{``You always have to \underline{propose challenges} to students}" (I4). When it is impossible to identify a text snippet that confirms the use of the teaching method during the course, the teaching method is considered not implemented.  

\subsubsection{Feedback from Participants (member-check)}

We used a follow-up survey with the DevOps educators interviewed by Fernandes \textit{et al.} \cite{fernandes:2022} to evaluate the interpretations of the teaching methods used in the educators' courses. Fernandes \textit{et al.} had not provided any member checking.
First, the survey instrument (online questionnaire) presents the aim of the work and defines what a teaching method is. The survey consisted of closed-ended and open questions. In closed-ended questions, educators indicate whether they agree or not with the teaching methods recognized in the interview transcripts. Answering these questions is mandatory.  
Open questions (optional) allow DevOps educators to communicate additional teaching methods that they often use during courses that were not recognized from the transcripts or to provide a link to course materials.

We sent the 14 personalized member-check surveys in May 2022 through their emails. After two weeks, we sent another email to reinforce the request for participants to contribute. We obtained feedback from 4 (28.5\%) participants I2, I5, I9, and I13, getting 
81.25\% strong agreement for the identified teaching methods. I2 strongly agrees about using the following teaching methods: \textsl{collaborative learning}, \textsl{educational support tool}, \textsl{seminars}, and \textsl{labs}. He was neutral about \textsl{lectures} and \textsl{review sessions}. I5 strongly agrees with using \textsl{labs} and project-based learning. I9 strongly agrees with using case studies, project-based learning, and collaborative learning. He also explained that many case studies are related to how organizations have evolved their DevOps practices and the challenges they had to overcome. According to I9, these case studies do not focus on the technical aspects, but on the organizational and people aspects. I13 strongly agrees about using \textsl{project-based learning}, \textsl{lecture}, \textsl{tutorial}, and \textsl{collaborative learning}. He was neutral about using \textsl{labs}.

\subsection{Linking Teaching Methods to Challenges}

We also linked the identified teaching methods to the challenges extracted from the interviews conducted by Fernandes \textit{et al.}~\cite{fernandes:2022}. They presented 558 association links between recommendations and challenges as a means of solving or mitigating challenges. According to them, a challenge is defined as a problem that makes it difficult to plan or run a DevOps course, while a recommendation is a real proposition to deal with a problem, easing the learning process. Some of their recommendations involve applying a teaching method such as \textit{``Use of \textbf{Problem-based Learning} (PBL) for the teaching of DevOps."} from I7. In this regard, we analyze the links between challenges and recommendations,  checking whether the recommendation is related to a teaching method.  In the case of any evidence, the researcher records the presence of a link between the teaching method and the challenge. Another researcher reviewer also checks the found links and discusses possible conflicting opinions, as explained at the beginning of the section.

We analyze the 558 links created by Fernandes \textit{et al.}~\cite{fernandes:2022}. We identified 87 links involving the same respondent about the challenge and the recommendation. We also identified links that involved different respondents about the challenge and the recommendation. Our study is restricted to the analysis of links that involve the same interviewee in the challenge and recommendations. These links were more consistent, as they occurred in the interviewee’s courses.

\section{Results}  \label{sec:results}

In this section, we present the answers to the research questions of our study.  We provide a set of artifacts~\cite{researchArtifact} that contain all the data collected and analyzed in this paper.

\subsection{RQ1. \textsc{What are the teaching methods and approaches in DevOps courses?}}

\begin{figure}[ht]
 \centering
  \includegraphics[width=0.49\textwidth]{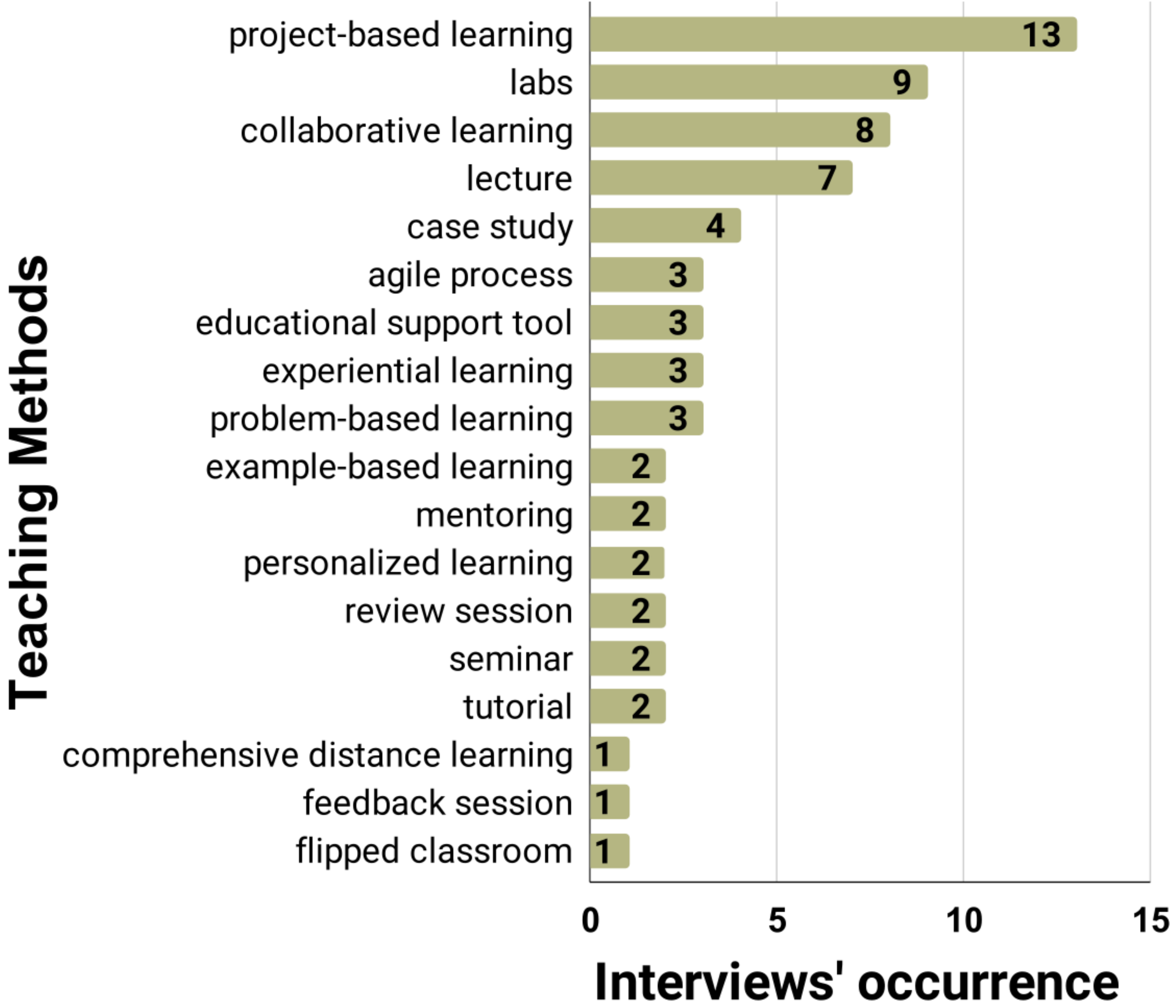}
  \caption{18 teaching methods distributed across 14 interviews.}
  \label{fig:interviewsResults}
\end{figure}

Figure~\ref{fig:interviewsResults} shows the distribution of the 18 teaching methods identified in the interview transcript. We can observe that the four most cited methods are \textsl{project-based learning} (13 interviewee citations), \textsl{labs} (9 interviewee citations), \textsl{collaborative learning} (8 interviewee citations), and \textsl{lecture} (7 interviewee citations). On the other hand, the four least cited methods are \textsl{comprehensive distance learning}, \textsl{feedback sessions}, \textsl{flipped classroom}, and \textsl{tutorial} with only one interviewee citation each. We identified an average of 4 teaching methods by interview (maximum: 7; minimum: 2).

\begin{table*}[ht]
\centering
\caption{Teaching methods definitions}
\label{tab:teachingMethodsDefinition}
\begin{tabular}{|c|c|} 
\hline
\multicolumn{1}{|c}{\textbf{{Teaching Methods}}} & \multicolumn{1}{c|}{\textbf{{Definitions}}}     \\ 
\hline
\textsc{Project-based~learning}                        & \begin{tabular}[c]{@{}c@{}}{It focuses on a project in which the students work}\\{ on a concrete task \cite{indiramma:2014}.}\end{tabular}             \\ 
\cline{2-2}
\textsc{Collaborative learning}                        & \begin{tabular}[c]{@{}c@{}}{The students work collaboratively exchanging information and }\\{resolving tasks. The teacher is the active partner~not just a repository of }\\{the information such as traditional education~\cite{pivec:2003}.}\end{tabular}      \\ 
\cline{2-2}
\textsc{Flipped classroom}                             & \begin{tabular}[c]{@{}c@{}}{Activities traditionally conducted in the classroom become }\\{home activities (vice versa)~\cite{akccayir:2018}.}\end{tabular}   \\ 
\cline{2-2}
\textsc{Seminar}                                       & \begin{tabular}[c]{@{}c@{}}{Students work in groups to discuss assigned questions and issues under }\\{ the guidance of teachers. As an outcome of the seminars, the students}\\{ give presentations or write an essay on their particular topic \cite{zeng:2020, kuhrmann:2012}.}\end{tabular}        \\ 
\cline{2-2}
\textsc{Lecture}                                       & \begin{tabular}[c]{@{}c@{}}{It is the traditional teaching method in educational institutions}\\{in which the lecturer directly instructs the students \cite{omatseye:2007, kuhrmann:2012}.}\end{tabular}  \\ 
\cline{2-2}
\textsc{Labs}                                          & \begin{tabular}[c]{@{}c@{}}{It involves accomplishing practical tasks exploring a computer science topic}\\  {usually conducted in dedicated rooms equipped with computers for each student.\cite{hazzan:2020}.} \end{tabular}  \\ 
\cline{2-2}
\textsc{Review session}                                & \begin{tabular}[c]{@{}c@{}}{In the review session, the answer (definition) is given and the student }\\{must come up with the question, a process that is the reverse }\\{of their study process \cite{harder:1992}.}\end{tabular}     \\ 
\cline{2-2}
\textsc{Problem-based~learning}                        & \begin{tabular}[c]{@{}c@{}}{It is a way of constructing and teaching courses using problems~}\\{as the motivation and focus for students’ activity \cite{richardson:2009}.}\end{tabular}  \\ 
\cline{2-2}
\textsc{Personalized learning}                         & \begin{tabular}[c]{@{}c@{}}{It describes various instructional approaches aimed at}\\{ meeting the learning needs of individuals~\cite{akyuz:2020}}\end{tabular}  \\ 
\cline{2-2}
\textsc{Experiential learning}                         & \begin{tabular}[c]{@{}c@{}}{It is the process of learning from concrete experience and reflecting} \\{ about into real-world situations \cite{kolb:1984}.} \end{tabular}  \\ 
\cline{2-2}
\textsc{Educational support tool}                      & \begin{tabular}[c]{@{}c@{}}{A tool or integrated environment used to support}\\{ the teaching or learning \cite{shahid:2019}.} \end{tabular}        \\ 
\cline{2-2}
\textsc{Comprehensive distance learning}        &   \begin{tabular}[c]{@{}c@{}}{It is an instructional model consciously selected in advance, with time to}\\{plan and make preparations to better ensure quality and accessibility}\\{ of the learning experience for all students \cite{maletz:2021}.}\end{tabular}     \\ 
\cline{2-2}
\textsc{Agile process}                                 & \begin{tabular}[c]{@{}c@{}}{Use of Agile activities during course execution like }\\{sprints and scrum planning \cite{salza:2019}.}\end{tabular}     \\ 
\cline{2-2}
\textsc{Case study}                                    & \begin{tabular}[c]{@{}c@{}}{Practical examples encouraging students to integrate knowledge}\\{~from class with real life \cite{krusche:2017}.}\end{tabular}     \\ 
\cline{2-2}
\textsc{Feedback session}                              & \begin{tabular}[c]{@{}c@{}}{The educator gives general feedback on the set of received}\\{ assignments focusing on the aspects that most people did right,}\\{ and the problems that were found  \cite{isomottonen:2013}.}\end{tabular} \\ 
\cline{2-2}
\textsc{Mentoring}                              & \begin{tabular}[c]{@{}c@{}}{It is an approach to improving teaching by adopting a mentor to help the}\\{educator during educational activities. A mentor helps the team organize their work}\\{ and tracks if the team’s planned didactic results are being achieved \cite{dowdall:2021}.}\end{tabular} \\ 
\cline{2-2}
\textsc{Example-based learning}                              & \begin{tabular}[c]{@{}c@{}}{It's based on providing worked examples that illustrate a written}\\{account of how a problem should be  or can be solved\cite{van:2010}.}\end{tabular} \\ 
\cline{2-2}
\textsc{Tutorial}                                      & \begin{tabular}[c]{@{}c@{}}{A video or documentation with the purpose of to introduce the more}\\{general~reader to the theoretical or technical concepts}\\{as well as configurations~steps \cite{lokkila:2016}.}\end{tabular} \\
\hline
\end{tabular}
\end{table*}

The rest of this subsection presents information about the identified teaching methods. We highlight how educators implemented those teaching methods in their respective courses.

{{\textbf{Project-based Learning}.}} This teaching method focuses on a project in which students work on a concrete task \cite{indiramma:2014}. I5 motivates students to implement their projects from scratch. The students incrementally include DevOps practices (e.g., Automated Build, Automated Testing, and Continuous Integration) in the same project. However, the educator also considers giving a starter project to reduce the difficulties faced by students during the initial implementation of the project. The course project can also involve an open-source project in which students should have to improve by adding DevOps practices. In this context, I11 gives students the initial code to work on.

I1 proposes to develop a software project in incremental parts to teach students to deal with complexity. Different modules related to DevOps are required during incremental development, such as adding containerization with a unique container, then a cluster of containers.

I11 argues that students should work on innovative projects. I7 allows students to develop projects related to other areas of computer science, such as artificial intelligence, which are more attractive to them. I13 motivates students to contribute to open-source projects with more than 100 stars. I14 also encourages students to work on industrial projects that enable them to understand how to work in realistic scenarios and why mindset is essential.

I6 mentions the importance of planning the project requirements on which the students will work. This involves considering the programming language/frameworks and the associated tools. For example, they have adopted the Java language because the teaching staff has more confidence in this programming language. I6's course also adopts popular and recommended DevOps tools. Another requirement is to use tools that students can install and run on their computers. In this sense, I10 delimits a set of tools/technologies that students could adopt in their projects, giving freedom only to the functional aspect of the projects. On the other hand, I13 gives students total freedom to choose the tools used in the project. I9 gives examples of projects to students to help them understand how to meet the project requirement. I11 summarizes the project requirements in a document.

I1 and I3 also argue the importance of a project-based approach as an alternative to course evaluation in the face of traditional exams. On the other hand, I4 has projects throughout the course, but does not evaluate them. I4 employs a project-based approach in which students are observed and the teacher tries to identify their difficulties and how they handle them. But I4 also uses the Net Promoter Score (NPS) to get feedback on the course from the students.

{{\textbf{Labs}.}} I3 explains that there are many ways to demonstrate a DevOps concept, supported by a large tools ecosystem. However, I5 reminds students that the course focuses on practicing the concepts with many tools, but not mastering these tools. In this context, I3 utilizes relevant free tools/technologies during labs. They use free services on public clouds. At the same time, I12 adopts open-source tools during labs. This enables students to practice in a realistic environment.

I8 automatizes the creation of the labs using snapshots. It facilitates the operation of the labs, mainly in large classes, where it is challenging to have personalized attendance. The educator also highlights the importance of observing the update of tools used in the labs, since new versions can break existing configurations. I10 and I14 faced the same problem. I11 and I12 report that the help of teaching assistants with proficiency in the tools has contributed to reducing the effort to prepare the labs.

\textbf{Collaborative learning}. Collaboration is one of the essential DevOps foundations. DevOps promotes collaboration between development and operation teams. I8 reinforces the importance of collaboration, as it is very common in the industry. I2 uses a Continuous Integration (CI) tool to track student collaboration in groups. I6 observes whether students work together or if each team member works alone. However, they think that it is not an efficient approach, as it takes a long time for the educator.

I6 uses problems and projects in groups to encourage student collaboration. I13, I11, and I8 use only team projects to teach how to work collaboratively in a team. I11 highlights that working in collaboration allows students to share problems with each other. I10 reinforces the need to be concerned about the differences between the environments of the team members. For example, differences in software versions could break the project build/pipeline.

I8 utilizes pair programming, where both students come up with the answers. I8 also creates a collaborative slack workspace. In this workspace, the educator and the students can send messages to each other. I8 explains to students that it is their responsibility as team members to keep their team motivated.

\textbf{Lectures}. I2 shares recorded lectures with students, complemented by the virtual classroom. They use these virtual classrooms to answer and discuss student questions. I4 improves the dynamic of lectures by using strategies such as changing their voice and interacting with the students. I2 explains that it is easier for students not to focus in the classroom. I13 invites industrial guests to give part of the lectures to motivate and present students with real scenarios.

I11 utilizes lectures to teach and explain DevOps concepts and related challenges. They also introduce essential DevOps tools, such as Jenkins, a CI tool. I12 uses the lecture to share the experience and to detail specific DevOps concepts. They teach Agile and how to operate the Kanban board - DevOps concepts according to Leite \textit{et al.} \cite{leite:2019}. I12 also teaches containerization techniques through lectures.

\textbf{Other teaching methods}. I12 uses case studies from relevant companies such as Google, Meta, and Netflix to present a couple of industrial experiences in the classroom. I13 motivates the students to use the Katacoda\footnote{\url{https://katacoda.com}} platform to share tutorials on the tools used throughout the course. I4 implements \textsl{personalized learning} through personalized support teams that support student challenges when setting up the tools on their machines.

\textbf{Combining teaching methods}. Table \ref{tab:strategiesUsed} shows the teaching methods distributed across the 13 courses. 
We categorized the teaching method occurrences in the interviews' transcripts into three levels: 
$\CIRCLE$ used in class, $\LEFTcircle$ just as a recommendation, and $\Circle$ not identified. Each interview corresponds to a unique course, except I10 and I14 related to the 10th course. The black-gray color indicates that we found evidence of the teaching method adopted in the course. The light gray color indicates that the method was mentioned and that we could not find evidence of its implementation. In this context, we understand that the interviewee only recommends adopting the teaching method throughout the course. In this sense, 17 (94.4\%) of the 18 identified teaching methods follow the implemented method category. As a highlight, \textsl{comprehensive distance learning} appears only in the category of recommended teaching methods. We also did not identify any use of teaching methods in the I6 interview, while I1 only implements \textsl{project-based learning}.

\begin{table*}[ht]
\centering
\caption{18 teaching methods distributed across the 13 courses}
\label{tab:strategiesUsed}
\begin{tabular}{|cccccccccccccc|} 
\hline
                                & \multicolumn{13}{c|}{\textbf{Courses}}                                                                                                                                  \\ 
\cline{2-14}
\textbf{Teaching Methods}       & ~1~        & ~2~        & ~3~        & ~4~        & 5          & ~6~        & ~7~        & ~8~        & ~9~        & 10         & 11         & 12         & 13          \\ 
\hline
\textsc{Project-based~learning}          & $\CIRCLE$ & $\Circle$ & $\CIRCLE$ & $\CIRCLE$ & $\CIRCLE$ & $\LEFTcircle$ & $\CIRCLE$ & $\CIRCLE$ & $\CIRCLE$ & $\CIRCLE$ & $\CIRCLE$ & $\Circle$ & $\CIRCLE$  \\
\textsc{Collaborative learning}          & $\Circle$ & $\CIRCLE$ & $\Circle$ & $\CIRCLE$ & $\Circle$ & $\LEFTcircle$ & $\Circle$ & $\CIRCLE$ & $\LEFTcircle$ & $\CIRCLE$ & $\CIRCLE$ & $\Circle$ & $\CIRCLE$  \\
\textsc{Seminar}                         & $\Circle$ & $\LEFTcircle$ & $\Circle$ & $\Circle$ & $\Circle$ & $\Circle$ & $\Circle$ & $\Circle$ & $\Circle$ & $\CIRCLE$ & $\Circle$ & $\Circle$ & $\Circle$  \\
\textsc{Lecture}                         & $\Circle$ & $\CIRCLE$ & $\Circle$ & $\LEFTcircle$ & $\Circle$ & $\Circle$ & $\Circle$ & $\Circle$ & $\Circle$ & $\CIRCLE$ & $\CIRCLE$ & $\CIRCLE$ & $\CIRCLE$  \\
\textsc{Labs}                            & $\Circle$ & $\CIRCLE$ & $\CIRCLE$ & $\Circle$ & $\CIRCLE$ & $\Circle$ & $\Circle$ & $\CIRCLE$ & $\Circle$ & $\CIRCLE$ & $\CIRCLE$ & $\CIRCLE$ & $\CIRCLE$  \\
\textsc{Review session}                  & $\Circle$ & $\CIRCLE$ & $\Circle$ & $\CIRCLE$ & $\Circle$ & $\Circle$ & $\Circle$ & $\Circle$ & $\Circle$ & $\Circle$ & $\Circle$ & $\Circle$ & $\Circle$  \\
\textsc{Problem-based~learning}          & $\Circle$ & $\Circle$ & $\Circle$ & $\LEFTcircle$ & $\Circle$ & $\Circle$ & $\CIRCLE$ & $\Circle$ & $\Circle$ & $\LEFTcircle$ & $\Circle$ & $\Circle$ & $\Circle$  \\
\textsc{Personalized learning}           & $\Circle$ & $\Circle$ & $\Circle$ & $\CIRCLE$ & $\Circle$ & $\Circle$ & $\Circle$ & $\CIRCLE$ & $\Circle$ & $\Circle$ & $\Circle$ & $\Circle$ & $\Circle$  \\
\textsc{Experiential learning}           & $\Circle$ & $\Circle$ & $\Circle$ & $\Circle$ & $\Circle$ & $\Circle$ & $\LEFTcircle$ & $\CIRCLE$ & $\Circle$ & $\Circle$ & $\Circle$ & $\Circle$ & $\Circle$  \\
\textsc{Educational support tool}        & $\LEFTcircle$ & $\CIRCLE$ & $\Circle$ & $\Circle$ & $\Circle$ & $\Circle$ & $\CIRCLE$ & $\Circle$ & $\Circle$ & $\Circle$ & $\Circle$ & $\Circle$ & $\Circle$  \\
\textsc{Comprehensive distance learning} & $\Circle$ & $\Circle$ & $\Circle$ & $\Circle$ & $\Circle$ & $\Circle$ & $\LEFTcircle$ & $\Circle$ & $\Circle$ & $\Circle$ & $\Circle$ & $\Circle$ & $\Circle$  \\
\textsc{Flipped classroom}               & $\Circle$ & $\Circle$ & $\Circle$ & $\Circle$ & $\Circle$ & $\Circle$ & $\CIRCLE$ & $\Circle$ & $\Circle$ & $\Circle$ & $\Circle$ & $\Circle$ & $\Circle$  \\
\textsc{Agile process}                   & $\Circle$ & $\Circle$ & $\Circle$ & $\Circle$ & $\Circle$ & $\Circle$ & $\CIRCLE$ & $\CIRCLE$ & $\Circle$ & $\Circle$ & $\Circle$ & $\CIRCLE$ & $\Circle$  \\
\textsc{Case study}                      & $\Circle$ & $\Circle$ & $\Circle$ & $\Circle$ & $\Circle$ & $\Circle$ & $\Circle$ & $\Circle$ & $\CIRCLE$ & $\CIRCLE$ & $\Circle$ & $\CIRCLE$ & $\Circle$  \\
\textsc{Feedback session}                & $\Circle$ & $\Circle$ & $\Circle$ & $\Circle$ & $\Circle$ & $\Circle$ & $\Circle$ & $\Circle$ & $\Circle$ & $\Circle$ & $\Circle$ & $\CIRCLE$ & $\Circle$  \\
\textsc{Tutorial}                        & $\Circle$ & $\Circle$ & $\Circle$ & $\Circle$ & $\Circle$ & $\Circle$ & $\Circle$ & $\Circle$ & $\Circle$ & $\Circle$ & $\Circle$ & $\Circle$ & $\CIRCLE$  \\ 
\textsc{Mentoring}                        & $\Circle$ & $\Circle$ & $\Circle$ & $\CIRCLE$ & $\Circle$ & $\Circle$ & $\Circle$ & $\Circle$ & $\Circle$ & $\Circle$ & $\CIRCLE$ & $\Circle$ & $\Circle$  \\ 
\textsc{Example-based learning}                        & $\Circle$ & $\Circle$ & $\CIRCLE$ & $\Circle$ & $\CIRCLE$ & $\Circle$ & $\Circle$ & $\Circle$ & $\Circle$ & $\Circle$ & $\Circle$ & $\Circle$ & $\Circle$  \\ 
\hline
\textbf{\textsc{TOTAL}}                  & 2          & 6          & 3          & 8          & 3          & 2          & 7          & 6          & 3          & 7          & 5          & 5          & 5           \\
\hline
\end{tabular}
\caption*{Key: $\CIRCLE$ =  used in class; $\LEFTcircle$ = just as a recommendation; $\Circle$ = not identified.}
\end{table*}

Analyzing Table \ref{tab:strategiesUsed}, it includes 13 combinations of implemented teaching methods together with teaching methods cited during interviews, without evidence of implementation. It also includes 12 combinations with only the teaching methods implemented during the class. For example, we identified \textsl{collaborative learning}, \textsl{mentoring}, \textsl{lecture}, \textsl{personalized learning}, \textsl{problem-based learning}, \textsl{project-based learning}, and \textsl{review session} during the interview with I4. We could identify evidence of implementation of \textsl{project-based learning}, \textsl{collaborative learning}, \textsl{review session}, \textsl{personalized learning}, and \textsl{mentoring} during their course.

\subsection{RQ2. \textsc{How can teaching methods address the challenges in DevOps courses?}}

\begin{figure*}[ht]
 \centering
  \includegraphics[width=0.93\textwidth]{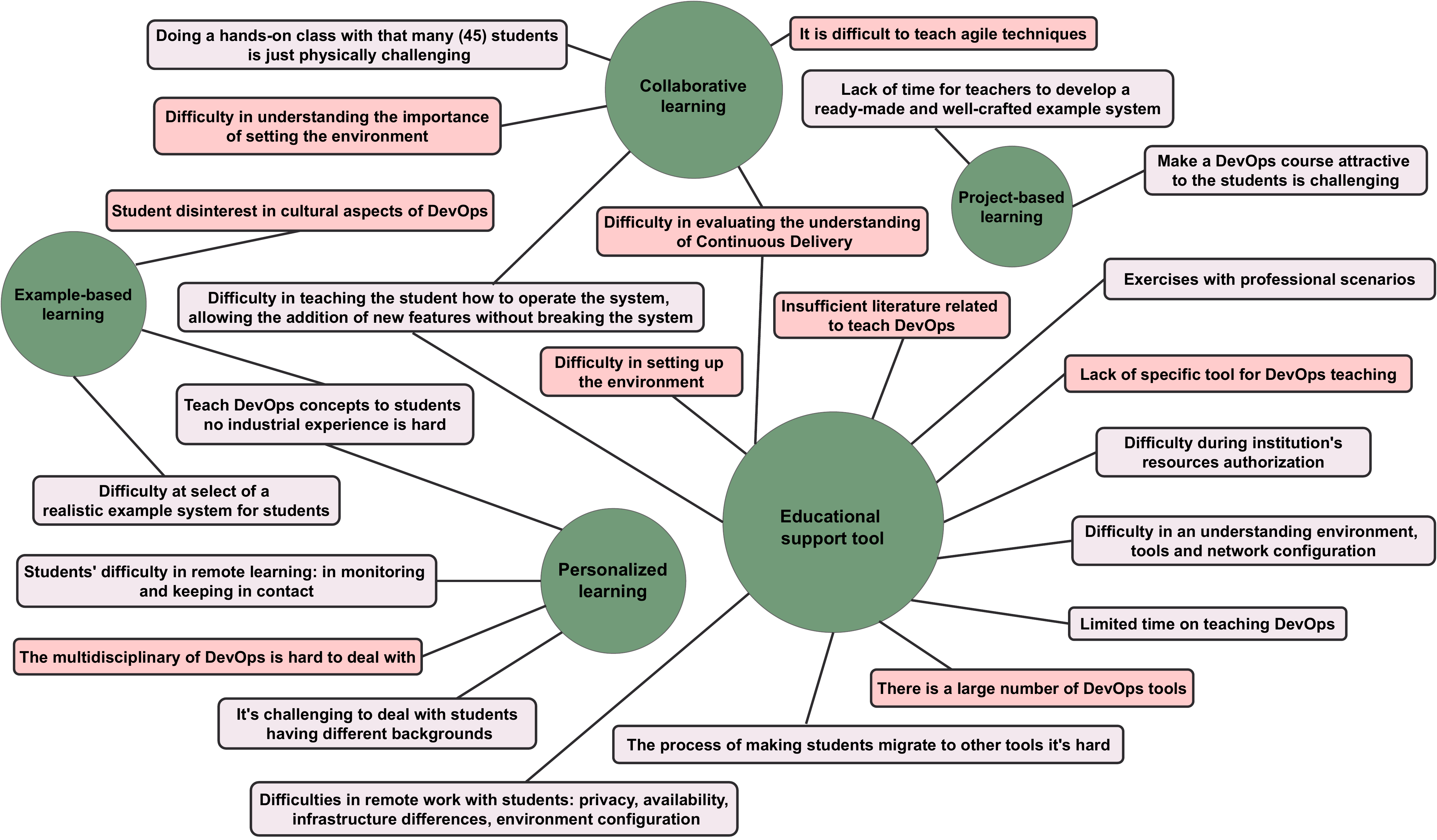}
   \caption{Most linked teaching methods to DevOps challenges.}
   Key: $\textcolor{teachingMethods}{\CIRCLE}$ = teaching method; $\textcolor{devopsSpecificChallenges}{\CIRCLE}$ = DevOps-specific challenge; $\textcolor{generalChallenges}{\CIRCLE}$ = general challenge. 
  \label{fig:mostLinkedTeachingMethods}
\end{figure*}

We identified 44 links between challenges and recommendations that involved the use of teaching methods. Figure \ref{fig:mostLinkedTeachingMethods} shows the most linked teaching methods that include \textsl{educational support tool}, \textsl{example-based learning}, \textsl{personalized learning}, \textsl{collaboration learning}, and \textsl{project-based learning}. We distinguish the DevOps-specific challenges using a $\textcolor{devopsSpecificChallenges}{\CIRCLE}$ background. Most challenges were linked to a unique teaching method, but we also have a few challenges linked to two teaching methods. For example, I3, I12, I13 and I14 shared the challenge that it is hard to teach DevOps concepts to students without industrial experience. Since DevOps motivation comes from solving conflicts between the development team and operations, students without industrial experience may not understand these conflicts. We identified \textsl{personalized learning} and \textsl{example-based learning} as teaching methods that address this challenge. I3 presents an example to students in a practical context from the initial stage to the final step. This worked example can complement students’ knowledge, providing a reference for them.

We could not identify links between challenges and recommendations for the following teaching methods:  \textsl{flipped classrooms}, \textsl{seminar},  and \textsl{tutorials}. Next, we describe the links found between the challenges and the recommendations that represent teaching methods.

{\textbf{Collaborative learning}.} I8 and I10 share the challenge of teaching Agile techniques. They also comment on the difficulty in hands-on classes with 45 or more students. In this sense, I8 uses a team-based student organization that can help mitigate these challenges. The experience of working as a team enables students to reflect on the importance of collaboration, an Agile principle \cite{diebold:2014}. At the same time, it contributes to helping educators keep students focused since they are organized in groups.

I2 comments on the difficulty in teaching the student how to operate the system, allowing the addition of new features without breaking the existing system functionalities. They comment on the challenge of understanding the Continuous Delivery concept by using and practicing infrastructure environment setup. According to I2: \textit{``The concept of Continuous Delivery [...] \textbf{The difficult} thing \textbf{is to put it into practice} [...] when they, as a team, need to release a certain functionality and ensure that it doesn't break the system''. For this reason,} I2 teaches the students social coding, a collaborative learning approach through the socialization of knowledge \cite{dabbish:2012}. In this context, social coding can motivate the students to learn interesting techniques and tools used by other student groups. This sharing of knowledge allows students to exchange information, produce ideas, simplify problems, and resolve tasks, as well as improve their understanding of DevOps practices.

{\textbf{Personalized learning}.} I3 and I7 share the perception that DevOps' multidisciplinary nature is difficult to manage because it usually involves coverage of diverse software engineering disciplines. I3, I12, I13, and I14 share that teaching DevOps concepts to students without industrial experience is hard. According to I3: \textit{``If the student is in a context where he has always been in the academic area, or he has never had practical experience with software development, [...] for the teacher, it becomes much more challenging to teach the DevOps concept to this student profile''}. For this purpose, I3 seeks to identify the most compatible DevOps scope for each class. They personalize the learning target according to the learning context of the students. Therefore, the educator should reflect and complement the initial level of knowledge of the students at the beginning of the course. For example, an educator can help students with little experience using Linux commands by sharing tutorials, creating labs, and encouraging discussions on DevOps topics.

I4, I5, I6, I8, and I10 commented that they had faced challenges in dealing with students of different backgrounds, making it difficult for students to collaborate in groups. According to I4: \textit{``One of the challenges is how do you teach people from these different backgrounds [...]  there is so much technology that comes together in DevOps, that the challenge is how do you get everyone up to speed on an even level? So that we can all move forward together and learn together. So, that's a big challenge''.} I4 and I5 deal with it by seeking to know the student's needs and limitations in advance. On the other hand, they suggest assuming that the students do not have much experience. I4 also avoids messing with student-specific problems, dealing with them in a personalized way at the right time. In other words, I4 alerts that the discussion of specific problems can eventually have a time limit; otherwise, the educator can lose the attention of the other students. I4 and I8 grade the students based on the student's learning journey and mistakes. They argue that the essential is how the students get there, understanding that every failure is a learning opportunity.

To mitigate this challenge of different backgrounds, I4 and I8 seek to develop a communication culture between students and educators, allowing a more comfortable learning environment for students. Another challenge of using this personalized approach is the difficulty in interacting with students when classes are remote, which I4, I8, and I12 faced. This interaction involves monitoring student progress, cleaning and discussing doubts, and keeping in touch.

\textbf{Problem-based learning}. I7 describes the difficulty in structuring the learning journey. I7 argues that the teaching plan should connect the covered DevOps subjects (for example, CI and Automated Tests). They use problem-based learning as a teaching method, starting from the problem and showing why people use it and what they are using. In this context,  Meyer \textit{et al.} \cite{meyer:2014} motivate the adoption of CI by facing the following main problem: \textit{``How do you verify whether someone's changes broke something in the code or whether the changes work in the larger context of the entire codebase?"}. Any educator could start teaching CI from this problem motivation and not only focus on teaching CI tools and their respective functionalities.

\textbf{Example-based learning}. I3, I4, I6, I9, I11, I12, and I14 emphasize that it is hard to show students that DevOps is not all about tooling (e.g., Docker and Jenkins). According to I3: \textit{``The student hopes to [...] learn that killer tool, which can help him in his professional career [...] he is usually more interested to know and learn the tools than understand the DevOps culture''}. There is little interest in learning the DevOps cultural aspects, such as collaboration. For this purpose, I3, I4, I11, and I12 use concrete examples during the teaching. These examples do not focus on a specific tool. Students should understand that there is no \textit{silver bullet} and that each problem can have a set of comprehensive solutions.

Another challenge mitigated by the \textsl{example-based learning} approach is that teaching DevOps concepts to students without industrial experience is difficult. Concrete examples can contribute to clarifying the student’s understanding. However, I5, I9, I11, and I14 highlight the difficulty in selecting realistic examples of systems for students. According to I5: \textit{``The challenge sometimes is finding a good open-source application, which is not too big also because you don't want the project to be too big. You \textbf{don't want it to be too small}, \textbf{but you don't want too big}''}. Simple systems can demotivate students. For this purpose, I5 plans to provide a ready-made sample system to students. As a benefit, the educator can have more confidence in supporting students during the course.

I11 and I12 share the challenge of creating an attractive DevOps course from the students’ perspectives. I11 worries that the class project should not be very small and must be challenging. Small projects such as toy projects may not be attractive to students since they are usually simple.

\textbf{Discussion}. I3, I4, I6, and I8 comment on the challenge of teaching DevOps culture. They also comment that there is no ready-made recipe (step-by-step instructions) to teach the DevOps mindset.  DevOps culture includes aspects such as continuous feedback and sharing of knowledge \cite{sanchez:2018}. I3 and I4 dealt with this challenge by promoting discussions of DevOps concepts and related issues. They use the students' difficulties, opinions, and experiences, pointing out solutions using DevOps. By doing this, the students could familiarize themselves more easily since the explanation is related to something they know well.

\textbf{Mentoring}. I4 shares difficulties in remote teaching. He highlights that students need to worry about setting up their computers while studying at home. It includes installing the necessary software. In this context, I4 created a specific support team to deal with infrastructure queries. This team helps students solve software infrastructure problems.

I7 mentions the need for great effort during the assessment of large classes. For this purpose, I7 highlights the importance of teaching assistants to help in the assessment process: \textit{``team of teaching assistants [...] If you don't have them, it's more difficult, you need to assess the projects alone. Take a class with 40 students, even if you divide it into teams, it's a lot of work...''}. Teaching assistants can help the leading educator.

\textbf{Agile processes}. I8 and I10 highlight the difficulty in teaching Agile techniques, as already mentioned. However, another approach to mitigate this challenge is to make the course Agile-oriented. I8 organizes the course in sprints with challenging incremental deliverables. In this context, we suggest the adoption of Agile techniques such as planning meetings and daily discussions \cite{diebold:2014} to make the student understand Agile processes and techniques through experimentation.

\textbf{Case study}. As mentioned above, I3, I4, I6, and I8 comment on the challenge of teaching DevOps culture. In this sense, I3 presents DevOps case studies in their course to students to help them understand cultural aspects. These case studies include the elimination of silos between development and operations teams; and the relationship between DevOps and the Site Reliability Engineering (SRE) professional.

\textbf{Project-based learning}. I11 reports that making a DevOps course attractive to students is a challenge: \textit{``You can make the lectures more interactive, but to make the lecture attractive 
\textbf{students have to be willing to interact}. [...] \textbf{Which is very difficult to do}''}. For this purpose, they argue that the project of the class should not be very small. Amorim \textit{et al.} \cite{amorim:2020} identified a lack of opportunity for student creativity, using small projects.

\textbf{Feedback Session}. We already commented that I8 and I10 had difficulty executing the hands-on classes with 45 or more students. In I8’s course, they have teaching assistants (TA) who help students with the management of technical questions related to the exercises. After the class, the TAs answer the remaining questions that are not handled during the class. They used an online shared document where all students could improve their understanding by reading the TAs' feedback. 

\textbf{Labs}. I1, I3, I5, I6, I8, I9, I10, and I11 mentioned the challenge of not having enough time in the course to teach DevOps. For example, I3 mentions: \textit{``the issue of laboratories [...] you always end up as a \textbf{matter of time versus class development}''}. This challenge occurs because DevOps incorporates knowledge of many practices. To illustrate, Leite \textit{et al.} \cite{leite:2019} developed a DevOps conceptual map that includes more than one hundred concepts. In this context, I1, I8, I10, and I11 seek to focus more classes on the practical part than DevOps’s theoretical part (e.g., DevOps mindset). They recommend that the courses have a practical part that occupies at least 80\% of the classes. Then, the understanding of the students can be improved through experimentation.

\textbf{Lecture}. I12 alerts about the gap between what the industry wants from students about DevOps and what the university teaches: \textit{``We hear from our industrial partners and from industry in general there's this \textbf{huge gap between what the industry needs and what university provides}''}. I12 seeks to mitigate this by carefully selecting industrial speakers to share their experiences with students. According to I12, inviting these experts allows the student to face real problems and scenarios. This enables students to reflect on the importance of DevOps practice in the industry.

\textbf{Educational support tool}. I5 comments that many students have difficulty understanding the practice of operations (for example, network configuration). These students have considerable proficiency in software development practices, but low proficiency in Linux commands and operations practices. They may also face problems configuring their machines and installing infrastructure software since they are not used to practicing it. In this context, I5 provides an initial configuration of the environment for students. They use a virtual machine snapshot that contains the required course software. It includes a containerization tool (e.g., Docker), CI/CD server (e.g., Jenkins), and artifact repository (e.g., Artifactory).

I2, I5, and I11 mentioned that the preparation of the course is challenging, since there is insufficient literature on teaching DevOps. For example, there is insufficient material to help DevOps educators prepare their courses. I2 also highlights the difficulties in assessing the student's understanding of continuous delivery techniques. To mitigate this challenge, they use an integrated environment in which students put their solutions to the exercises.

On the other hand, I2 exposes the challenge of making clear to students the importance of having a more realistic perspective of production. To address it, they adopt industrial tools in the curriculum. Students can learn DevOps concepts by practicing with the same tools used in companies such as Google and Microsoft, facing and reflecting on real problems. This approach facilitates the preparation of students for the industry. 

However, I2 commented that they faced difficulties in requiring students to adopt course tools. Many students are resistant to adopting these tools. I2 explains that students usually do not feel comfortable using new tools. To mitigate it, the educator explains to the students that they have a good grasp of the features of the tool and that the students will have good support for any questions. On the other hand, many educators (I2, I3, I5, I6, I7, I8, and I11) mention a related challenge: the educator (and teaching assistants) should be able to manage tools for each step of the DevOps pipeline taught in their courses.

Finally, I1 mentioned the difficulty of holding an institution's resources authorization. According to I1: \textit{``challenges are basically how do you rebuild an enterprise environment into a \textbf{university environment} that \textbf{is} much \textbf{more restrictive} and doesn't have enough machines for them"}. This restriction includes, for example, the inability to install tools and network limitations to create near-real infrastructure as related challenges. For this reason, I1 uses cloud providers' services as infrastructure during the course. Although these cloud services usually require payment, I1 recommends that students use the students' free plan. I1 defends this approach by: \textit{``I recommend [...] \textbf{Moving all teaching to a cloud}. [...] contact AWS. They have a student program, or Google, with Ali Baba, Azure, and IBM Cloud"}.

\section{Discussion}  \label{sec:discussion}

We discuss our results by comparing them with the findings presented by Ferino \textit{et al.} \cite{ferino:2021} and in terms of the implications of the results for educators and researchers.

\subsection{Comparing Teaching Methods From Interviews and Literature} 

We compared our results with a systematic literature review that investigates existing DevOps teaching methods conducted by Ferino \textit{et al.} \cite{ferino:2021}. Their study selects a total of 18 papers reporting DevOps teaching experiences until the beginning of 2020. In this context, we first update Ferino \textit{et al.}'s results \cite{ferino:2021} by following their SLR methodology, including more 15 papers published between 2020 and 2021.   We finished with a total of 33 papers, referenced in our online appendix \cite{researchArtifact}.

\begin{figure}[ht]
 \centering
  \includegraphics[width=0.48\textwidth]{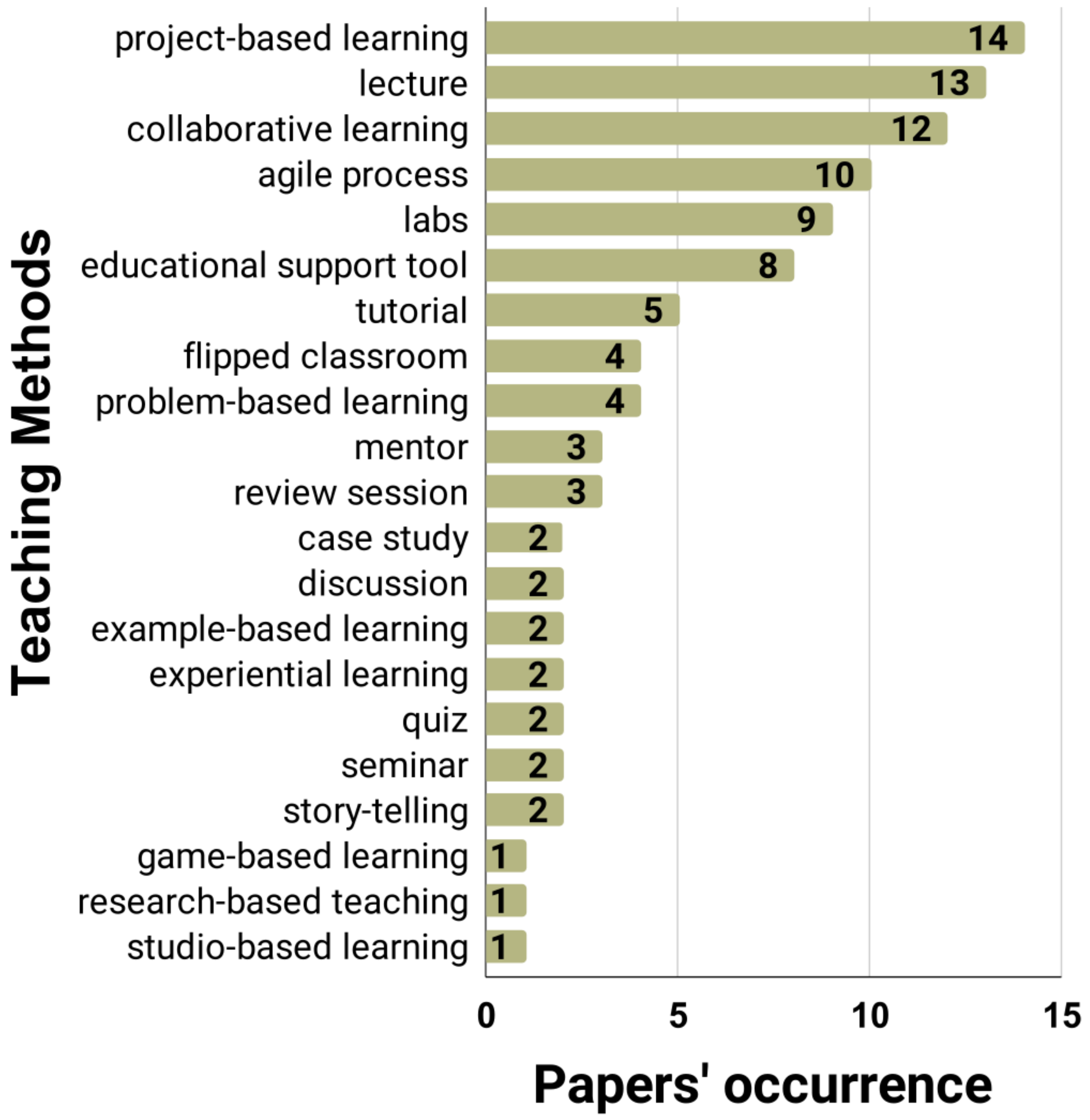}
  \caption{Distribution of 21 teaching methods by 33 papers.}
  \label{fig:rslResults}
\end{figure}

Figure \ref{fig:rslResults} presents the 21 teaching methods identified in the SLR considering our update. The four most cited methods are \textsl{project-based learning} (14 paper citations), \textsl{lecture} (13 paper citations), \textsl{collaborative learning} (12 paper citations), and \textsl{Agile process} (10 paper citations). On the other hand, the three least cited methods are \textsl{studio-based learning}, \textsl{game-based learning}, and \textsl{research-based teaching} with only one paper citation each. They refer to unusual teaching methods. \textsl{Game-based learning} refers to the use of computer games and games in general for educational purposes \cite{pivec:2003}. \textsl{Studio-based learning} focuses on student activity that involves interactions with peers and instructors. Studios are experiential learning opportunities based on projects / products. In studios, students discover for themselves what content is needed, what questions need to be answered, and what the answers are \cite{trede:2021}. \textsl{Research-based teaching} refers to students actively exploring and solving problems using scientific research approaches under the guidance of educators \cite{ye:2017}. Finally, we identified an average of 3 teaching methods by paper (maximum: 8; minimum: 1).

Our comparison identified a set of similar teaching methods from both sources: interviews and papers. As a highlight, both emphasize that \textsl{project-based learning} and \textsl{collaborative learning} are the most recurrent teaching methods. Most of the current DevOps courses focus on letting students work together on projects. This is in line with industry needs: professionals who have both hard and soft skills \cite{rabelo:2022}. Hard skills are mastered through working on projects, while soft skills proficiency is developed through collaborative work. Rabelo \textit{et al.} \cite{rabelo:2022}  identified \textbf{\textit{teamwork}} as the most in-demand non-technical skill in software development jobs today. Shih \textit{et al.} \cite{shih:2014} found that IT professionals often have introverted personalities. They show that this can be a barrier to understanding the needs and requirements of the customer \cite{rabelo:2022}. Collaborative work helps students improve their communication skills. For this purpose, DevOps courses today have a great impact on the formation of students as IT professionals.

We identified only four studies \cite{eddy:2017_pp, perez:2021_pp, abirami:2021_pp, okolica:2020_pp} investigating how effective these teaching methods are in the context of DevOps Education. Many studies only have student feedback surveys as a teaching method evaluation approach \cite{christensen:2018_pp, kuusinen:2019_pp, krusche:2014_pp, christensen:2016_pp, bobrov:2019_pp, alves:2021_pp, mirhosseini:2019_pp, kousa:2020_pp, rahman:2021_pp, hills:2020_pp}. The absence of evaluation studies on the adoption of teaching methods in DevOps education represents an opportunity for researchers.

Eddy \textit{et al.} \cite{eddy:2017_pp} present an empirical study that evaluates whether the use of a Continuous Delivery (CD) pipeline in a laboratory environment helps students better understand CD-related concepts. They found that the pipeline and activity were useful for students' understanding of the concepts. Okolica \textit{et al.} \cite{okolica:2020_pp} conducted a pilot study on \textsl{game-based learning} in a graduate-level course, where the lesson objectives focus on DevSecOps. The game was tested at the end of the course. Their results show that the students consider the game relevant to illustrate the topic.  

Perez \textit{et al.} \cite{perez:2021_pp} analyze whether the use of the \textsl{research-based teaching} (RBT) experience was effective for student learning. They also analyze whether inter-coder agreement (ICA) can be applied to verify student knowledge acquisition. As a result, the students showed homogeneous academic performance compared to the traditional-learning student group. However, they argue that the DevOps RBT experience has an inherent advantage in providing motivational learning as students acquire knowledge from their research. Abirami \textit{et al.} \cite{abirami:2021_pp} present an experimental study of applying various teaching methods in a DevOps course. They focused on mixing the following learning approaches: \textsl{discussions}, \textsl{seminars}, and \textsl{quizzes}. Unfortunately, we only identified this study working on blended learning in DevOps Education. They used the \textit{logistic regression model} to measure the impact of the blended learning approach on student learning. They identified that when these teaching methods are used, students have a significant impact on increasing their learning outcomes. This approach emerges as a starting point for new research that measures the impact of integrated teaching methods on student learning. Next, we describe the similarities and differences in applying the teaching methods reported in the interviews and literature.

\textbf{Similarities}. We confirm that 15 (83.3\%) of the 18 teaching methods derived from the interviews were also identified in the literature review. \textsl{Lectures} are highlighted in both sources. This is a simple, fast, and inexpensive method to present the vast issues to many groups of learners \cite{sadeghi:2014, golafrooz:2010}. In remote teaching, \textsl{lectures} is highlighted as a low-cost approach. In this context, Hobeck \textit{et al.} \cite{hobeck:2021_pp} mentioned that \textsl{flipped classroom} facilitated the transition to fully remote teaching during the COVID pandemic. They reported that the pre-class activities were unchanged since the lectures were already recorded on video, although the in-class activities changed significantly.

\textbf{Differences}. Our analysis identified that the following teaching methods were explored only in the papers of the SLR but not in the interviews of existing DevOps courses: \textsl{storytelling}, \textsl{game-based learning}, \textsl{research-based teaching}, \textsl{studio-based learning} and \textsl{quizzes}. We understand that educators could not be familiar or secure enough about how to implement these approaches, since DevOps is a new area. On the other hand, \textsl{personalized learning}, \textsl{comprehensive distance learning}, and \textsl{feedback session} appear only in the interviews.

Christensen \textit{et al.} \cite{christensen:2018_pp, christensen:2016_pp} use \textsl{storytelling} to design the learning context of DevOps topics in an existing undergraduate course. In this context, Rao \textit{et al.} \cite{rao:2006} emphasize the potential of \textsl{storytelling} as a tool to improve \textsl{lectures}, engaging the student through stories. Ouhbi \textit{et al.} \cite{ouhbi:2021} conducted an empirical study exploring the effects of combining \textsl{storytelling} with \textsl{lectures}. They identified a great impact on female students, increasing their interest in pursuing a career in software engineering. Therefore, this approach can be adopted as a strategy to engage female students, increasing female participation in the area of software engineering.

We also identified interesting recommendations on how to apply teaching methods. For example, Krusche \textit{et al.} \cite{krusche:2014_pp} implement \textsl{project-based learning} adopting projects with real customers in the industry. The students implemented Continuous Delivery pipelines on customer projects, enabling fast customer feedback. Wei \textit{et al.} \cite{wei:2019_pp} and Alves \textit{et al.} \cite{alves:2021_pp} motivate the students to work on open-source projects. This enables students to collaborate using DevOps tools. They represent projects with potential customers without a roadmap, similar to real software projects. Alves \textit{et al.} argue that project restrictions are important in limiting the scope of the project. In this course, the students present a set of project themes, but the students can also work on other project themes, as long as they are approved. However, Wei \textit{et al.} let the students be free to use whatever technology they choose.

\subsection{Lessons Learned for Educators} 

Next, we discuss the lessons learned, focused on improving the teaching experience.

\textbf{Teaching DevOps using Practical Approaches}. Similar to Grotta \textit{et al.} \cite{grotta:2022}, our findings show that DevOps concepts should be taught using practical teaching methods. Teaching methods such as \textsl{project-based} and \textsl{problem-based learning} appear to be good choices. Using these approaches, a DevOps course can contribute to improving the problem-solving skills of students. Moreover, \textsl{project-based learning} and \textsl{collaborative learning} emerged as a recurrent teaching method combination. We notice the potential of this approach to be implemented as a basis of the DevOps course.

\textbf{Collaborative Learning fits with the Culture of Collaboration Principle}. In the paper, we have mentioned the importance of promoting collaboration practices in DevOps courses. Adopting \textsl{collaborative learning} allow students to understand the challenges of working in a team (e.g., keeping group members motivated) and overcome collaboration difficulties with the support of educators.

\textbf{Establishing Industry-Academia Collaboration}. Industrial \textsl{mentoring} and industrial guest lectures appear as a potential opportunity for starting collaboration between industry and academia. Although there are challenges involving external collaboration, industrial practitioners are usually more proficient in technologies than academics. This approach enables the students to learn the technical skills needed for the industry in a realistic scenario. The industry can also supply the demand for professionals capable of facing their challenges.

\subsection{Implications to Researchers} 

Next, we discuss research opportunities for new studies.

\textbf{Recommending New Teaching Methods}. Trede \textit{et al.} \cite{trede:2021}  suggest that more innovative learning and teaching approaches in engineering education can help prepare students for the future world of work. Our study identified a set of teaching methods. However, we urge researchers to seek to increase this set by adopting and combining different teaching methods. We recommend starting research with evaluated teaching methods in software engineering and other areas of computer science. We also recommend focusing on more active and practice-oriented approaches, as DevOps seems to fit these strategies.

{\textbf{How DevOps Courses Impact Students' Non-Technical Skills}.} Our study identified an expressive adoption of collaborative learning in DevOps courses. Trede \textit{et al.} \cite{trede:2021} identified that this teaching method is closely related to teamwork and leadership. These non-technical skills are very important in the industry. Thus, we reinforce the importance of teaching collaboration concepts in the DevOps courses as part of the computer science and/or software engineering academic curricula.

\textbf{More Adequate Teaching Approaches for Training in Industry}.  We identified two papers \cite{bobrov:2019_pp, mazzara:2018_pp} that focus on teaching DevOps in an industrial context. Both studies did not conduct empirical studies on the integration of teaching methods. Teaching in an industrial context has hard time constraints, with training duration no longer than a few weeks. For this purpose, an interesting research topic is to understand what combination of teaching methods will more effectively contribute to teaching DevOps concepts to employees of a software company.

\section{Threats to Validity} \label{sec:threatsToValidity}

In this section, we discuss threats to the validity of this study in the context of qualitative research~\cite{fernandes:2022, larios:2020, guba:1981, korstjens:2018}.

{\textbf{Credibility}.} This validity refers to whether the study findings are correctly drawn from the original data. We employed several approaches to ensure credibility: (a) we always had a second researcher reviewing the data analyzed by the first researcher during the execution of the study; (b) we carried out member checking on the teaching methods with the participants, allowing them to complement our analysis with teaching methods not previously identified; (c) we compared our findings from interviews with Ferino \textit{et al.} \cite{ferino:2021}'s findings from SLR, including more studies by following SLR's methodology.

{\textbf{Transferability}.} It is related to the extent to which our results can be transferred to other contexts (generalization). Our study extends Fernandes \textit{et al.} \cite{fernandes:2022}'s transferability limitation since we analyzed their interviews' transcripts. They selected 14 DevOps educators from three continents (North America, South America, and Europe) focusing on diversity, covering academia and industry.

{\textbf{Confirmability}.} It refers to the degree to which other researchers can verify the findings. We show the evidence for each identified teaching method by quoting participants. We share our data sheets including the teaching methods and links between the teaching methods and the challenges in our online research artifact~\cite{researchArtifact}.

\section{Conclusion}   \label{sec:conclusion}

Our study aims at recognizing teaching methods used in the planning and execution of DevOps courses. We identified 18 teaching methods from interviews conducted by Fernandes \textit{et al.}\cite{fernandes:2022}. In summary, our study revealed that educators should favor practical methods to teach DevOps. Indeed, \textit{project-based learning} was the most recurrent teaching method reported in the interviews. Also, as observed in the literature review, most DevOps courses focus on qualifying students to work together. Likewise, other teaching methods (e.g., \textit{problem-based learning}) appear to be sound alternatives for improving students' problem-solving skills.

Our work also introduced 44 links between teaching methods and DevOps education challenges. \textsl{Educational support tool}, \textsl{collaborative learning}, \textsl{personalized learning}, \textsl{example-based learning}, and \textsl{project-based learning} are the most used teaching methods by educators in their classes to mitigate the challenges. As reported in this work, there is a recurrent challenge in teaching DevOps culture.  The promotion of discussion enables educators to use students' experiences to explain how DevOps could handle the issues faced.  The students also face difficulty in implementing Continuous Delivery practices, which is an essential DevOps concept. A collaborative approach using social coding appears as a mitigation strategy. The students can improve their skills by learning interesting techniques from other students.

Finally, we also examined research opportunities — e.g., the recommendation of new teaching methods and learning considering non-technical skills in DevOps courses. We aim to explore the studied teaching methods and approaches for training DevOps professionals in the industry as future work.

\vspace{0.4cm}

\textbf{Acknowledgements.} We thank all the educators who contributed to our study. This work is partially supported by INES (www.ines.org.br), CNPq grant 465614/2014-0, CAPES grant 88887.136410/2017-00, FACEPE grants APQ-0399-1.03/17, PRONEX APQ/ 0388-1.03/14, and IFRN.

\bibliographystyle{IEEEtran}
\bibliography{IEEEabrv,references}

\end{document}